\documentclass[useAMS,usenatbib,fleqn]{mn2e}

\usepackage{mathptmx}
\usepackage{natbib}
\usepackage[english]{babel}
\usepackage[utf8]{inputenc}
\usepackage{amsmath}
\usepackage{amssymb}
\usepackage{graphicx}
\usepackage{textcomp}
\usepackage[colorinlistoftodos]{todonotes}

\newcommand{\kms}{km\,s$^{-1}$}

\def \hii {H\,{\textsc {ii}}}
\def \nii {[N\,{\textsc {ii}}]}
\def \oi {[O\,{\textsc {i}}]}
\def \oii {[O\,{\textsc {ii}}]}
\def \oiii {[O\,{\textsc {iii}}]}
\def \sii {[S\,{\textsc {ii}}]}

\def \sifa {1}
\def \caastro {2}
\def \aao {3}
\def \hawaii {4}
\def \anu {5}
\def \macquariephysics {{6}}

\title[The nature of kinematically offset AGN]{The SAMI Galaxy Survey: Unveiling the nature of kinematically offset active galactic nuclei}

\author[J.\,T.~Allen et al.]
{\parbox{\textwidth}{\raggedright J.\,T.~Allen$^{\sifa,\caastro}$\thanks{j.allen@physics.usyd.edu.au},
A.\,L.~Schaefer$^{\sifa,\caastro,\aao}$,
N.~Scott$^{\sifa,\caastro}$,
L.\,M.\,R.~Fogarty$^{\sifa,\caastro}$,
I.-T.~Ho$^\hawaii$,
A.\,M.~Medling$^\anu$,
S.\,K.~Leslie$^{\anu,\caastro}$,
J.~Bland-Hawthorn$^{\sifa,\caastro}$,
J.\,J.~Bryant$^{\sifa,\caastro,\aao}$,
S.\,M.~Croom$^{\sifa,\caastro}$,
M.~Goodwin$^{\aao}$,
A.\,W.~Green$^{\aao}$,
I.\,S.~Konstantopoulos$^{\aao,\caastro}$,
J.\,S.~Lawrence$^{\aao}$,
M.\,S.~Owers$^{\aao,\macquariephysics}$,
S.\,N.~Richards$^{\sifa,\caastro,\aao}$,
R.~Sharp$^{\anu}$
}\vspace{0.4cm}\\
\parbox{\textwidth}{$^{\sifa}$ Sydney Institute for Astronomy (SIfA), School of Physics, The University of Sydney, NSW 2006, Australia
\\$^{\caastro}$ ARC Centre of Excellence for All-sky Astrophysics (CAASTRO)
\\$^{\aao}$ Australian Astronomical Observatory, PO Box 915, North Ryde, NSW 1670, Australia
\\$^\hawaii$ Institute for Astronomy, University of Hawaii, 2680 Woodlawn Drive, Honolulu, HI 96822, USA
\\$^{\anu}$ Research School of Astronomy \& Astrophysics, Australian National University, Mount Stromlo Observatory, Cotter Road, Weston Creek, ACT 2611, Australia
\\$^\macquariephysics$ Department of Physics and Astronomy, Macquarie University, NSW 2109, Australia
}}

\date{\today}

\begin{document}
\maketitle


\begin{abstract}
We have observed two kinematically offset active galactic nuclei (AGN), whose ionised gas is at a different line-of-sight velocity to their host galaxies, with the SAMI integral field spectrograph (IFS). One of the galaxies shows gas kinematics very different to the stellar kinematics, indicating a recent merger or accretion event. We demonstrate that the star formation associated with this event was triggered within the last 100\,Myr. The other galaxy shows simple disc rotation in both gas and stellar kinematics, aligned with each other, but in the central region has signatures of an outflow driven by the AGN. Other than the outflow, neither galaxy shows any discontinuity in the ionised gas kinematics at the galaxy's centre. We conclude that in these two cases there is no direct evidence of the AGN being in a supermassive black hole binary system. Our study demonstrates that selecting kinematically offset AGN from single-fibre spectroscopy provides, by definition, samples of kinematically peculiar objects, but IFS or other data are required to determine their true nature.
\end{abstract}


\section{Introduction}

In the event of a merger between two massive galaxies, the supermassive black hole (SMBH) at the centre of each progenitor is expected to migrate towards the centre of the remnant \citep{Begelman80}. When the two SMBHs become gravitationally bound to each other in a binary orbit, they continue to spiral inwards through the effect of three-body scattering of nearby stars \citep{Quinlan96,Khan11,Khan13} until at small radii ($\lesssim0.01$\,pc) gravitational wave emission causes further shrinking of the orbit, eventually leading to the SMBHs coalescing \citep{Jaffe03}.

The remaining uncertainties in the mechanisms and timescales by which SMBHs coalesce, and their link to mergers which are a major driver of galaxy evolution, has led to a wide interest in identifying and characterising SMBH binary systems. However, identification of such systems has proved highly challenging. The only truly unambiguous cases are those in which two distinct sources are observed in X-ray or radio interferometric imaging, indicating the presence of two accreting SMBHs \citep{Owen85,Komossa03,Hudson06,Rodriguez06,Bianchi08}. These systems will remain the `gold standard' for SMBH binary identification, but the difficulty of obtaining the necessary observations, and the low redshift required to resolve the two sources, mean that significant sample sizes are currently beyond reach.

\citet{Graham15} and \citet{Liu15} each claim detection of a SMBH binary system in a quasar via periodic modulation of the quasar's brightness. The systems are intriguing and clearly warrant close investigation, but further confirmation is required before this method can be established as a viable technique for SMBH binary searches.

A further technique for identifying SMBH binary candidates relies on the orbital motion of binary active galactic nuclei (AGN). If two SMBHs are in orbit with each other, and each carries a reservoir of ionised gas travelling with the same velocity as the SMBH, then an optical spectrum of the system will, depending on orientation, show two sets of emission lines with different line-of-sight velocities, or double-peaked emission lines. Searches for such systems have produced many candidate SMBH binaries \citep{Smith10,Liu10a}, some of which have been confirmed by follow-up X-ray or radio imaging as described above \citep{Comerford11}. Alternatively, long-term spectroscopic monitoring of candidates can identify those whose line-of-sight velocities change over time, indicating a binary orbit \citep{Gaskell96,Eracleous12}, although other variation in the spectra can mimic orbital velocity changes \citep{Eracleous97}. Confirmation can also be found from optical/infrared imaging or, preferably, spatially resolved spectroscopy, if two distinct sources are seen corresponding to the two kinematic components \citep{Liu10b,Comerford11,McGurk11,Fu12,Villforth15}. However, samples selected purely on the basis of having double-peaked emission lines are heavily contaminated by other systems that produce such spectra: Keplerian rotation, outflows, or other kinematic disturbances can all result in double-peaked lines that mimic SMBH binaries \citep{Eracleous03,Shen11,Fu11}.

Recently, \citet[hereafter CG14]{Comerford14} proposed a related method to search for SMBH binaries in which only one SMBH is an AGN. In this case, a single set of emission lines is seen, but these lines are kinematically offset from the host galaxy redshift due to the same orbital motions that, in other cases, produce double-peaked emission lines. CG14 presented a sample of 351 such objects drawn from the Sloan Digital Sky Survey (SDSS; \citealt{York00}) spectroscopic observations of galaxies. AGN with asymmetric emission lines, double-peaked emission lines, or mismatched Balmer and forbidden lines, were not included in the sample, in order to reduce contamination by outflows and other kinematic features internal to the AGN. However, as noted by CG14, the sample may still be contaminated by many of the same processes that can produce double-peaked emission, including outflows and dust obscuration. As a result, follow-up observations are required to confirm the candidates as SMBH binary systems.

In this paper we examine in detail two of the kinematically offset AGN in the CG14 sample, using integral field spectroscopy (IFS) from the Sydney--AAO Multi-object Integral field spectrograph (SAMI) Galaxy Survey. The SAMI Galaxy Survey is an ongoing survey of $\sim$3400 low-redshift ($z$$<$0.12) galaxies with a wide range of masses and environments. For each galaxy, SAMI produces spatially-resolved spectra over a 15\arcsec{} field of view, covering the two wavelength ranges 3700--5700\,\AA{} and 6300--7400\,\AA{} at spectral resolutions of $\sim$1700 and $\sim$4500, respectively. The original SAMI instrument is described in \citet{Croom12}, recent upgrades and the sample selection for the survey are presented in \citet{Bryant15}, while \citet{Sharp15} and \citet{Allen15} give details of the data reduction and data products.

The IFS data from the SAMI Galaxy Survey provide a useful test of the proposed binary SMBH nature of the selected galaxies, primarily by supplying kinematic information about the stars and ionised gas throughout each galaxy. If the kinematic offset observed in the SDSS spectra is the result of the accreting SMBH being in a binary orbit with another, passive, SMBH, we would expect the velocity of the AGN to be decoupled from that of the gas at larger radii. Away from the centre of the galaxy, the gas should in essence be unaware of the binary nature of the SMBH. As the galaxies here have been selected to have a kinematic offset at the centre, the key observational test is that the kinematics away from the centre should be normal. In other words, a binary system should result in a discontinuity in the observed gas velocity, with a spatially unresolved region at the centre of the galaxy in which there is a kinematic offset, surrounded by rotating gas that is settled in the same gravitational potential as the stars.

Other potential causes of the kinematic offset would result in different spatially resolved signatures; some specific cases are described in Section~\ref{sec:discussion}. One important scenario is that of an outflow directed along the line of sight, which can produce the same signature of a central kinematic offset surrounded by more normal gas that is expected from a binary SMBH. To distinguish between these two scenarios, a careful analysis of the properties of the central gas is required. Typical signatures of outflows include broadened and/or asymmetric emission lines, increased electron density, and shock heating. The presence of some or all of these features would indicate an outflow rather than a binary SMBH origin for the observed offset. Given that outflows are very common among luminous AGN, while SMBH binaries are expected to be rare, all possible signatures of outflows must be eliminated in order to gain a secure identification of a binary system.

Throughout this paper we use a flat $\Lambda$CDM concordance cosmology: $H_0=70$\,km\,s$^{-1}$\,Mpc$^{-1}$, $\Omega_0=0.3$, $\Omega_\Lambda=0.7$. Unless otherwise specified, the emission-line transitions referred to are ${\rm\oi=\oi\,\lambda6300}$; ${\rm\oii=\oii\,\lambda3726+\oii\,\lambda3729}$; ${\rm\oiii=\oiii\,\lambda5007}$; ${\rm\nii=\nii\,\lambda6583}$; ${\rm\sii=\sii\,\lambda6716+\sii\,\lambda6731}$. Air wavelengths are used throughout. 

\section{Sample and data}

The objects examined here are drawn from the CG14 sample of kinematically offset Type 2 AGN. CG14 selected these AGN from the OSSY catalogue \citep{Oh11} of emission and absorption line properties, calculated from fits to galaxy spectra from the SDSS. The CG14 objects were selected to lie above the \citet{Kewley01} theoretical maximum starburst line in the \nii{} BPT diagram \citep{Baldwin81}, indicating the presence of an AGN, and to have a $>$3$\sigma$ difference between the emission and absorption line-of-sight velocities. AGN were rejected if their emission lines were asymmetric (weighted skewness $\ge$0.5) as such lines are typical of outflows.

\begin{figure}
\centering
\includegraphics[width=85mm]{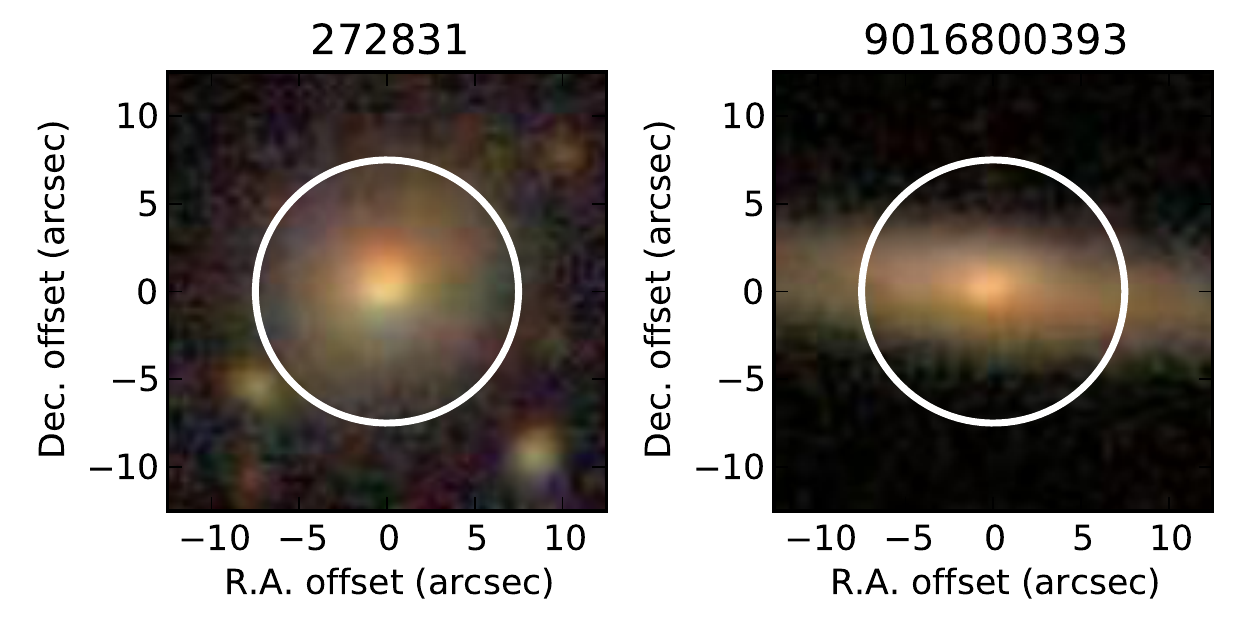} 
\caption{SDSS $gri$ composite images of the two galaxies examined in this paper, with IDs 272831 and 9016800393. The white circle in each panel shows the SAMI field of view.}
\label{fig:sdss}
\end{figure}

Two of the CG14 AGN have been observed as part of the SAMI Galaxy Survey, and in that context given the IDs 272831 and 9016800393. These two galaxies were not targeted for their AGN properties, but rather they met the redshift and stellar mass cuts used for all galaxies in the survey, described in \citet{Bryant15}. Some basic information about the two galaxies, and details of the observations, are given in Table~\ref{tab:observations}. The redshifts are measured from the SDSS spectra, while the stellar masses are calculated from the observed $g$ and $i$ magnitudes according to equation~3 in \citet{Bryant15}. Fig.~\ref{fig:sdss} shows SDSS $gri$ composite images of the two galaxies.

\begin{table*}
\centering
\caption{Catalogue and observational information about the two kinematically offset AGN and the SAMI Galaxy Survey observations.}
\label{tab:observations}
\begin{tabular}{rrr}
\hline
SAMI ID & 272831 & 9016800393 \\
\hline
SDSS name & J120401.97+012641.5 & J011333.05+002948.0 \\
R.A.\ (J2000 deg) & 181.00821 & 18.38774  \\
Dec.\ (J2000 deg) & 1.44487 & 0.49667  \\
Redshift & 0.08352 & 0.04404  \\
$\log_{10}$(Stellar mass / M$_\odot$) & 11.17 & 10.63  \\
OSSY weighted offset (\kms) & $67\pm13$ & $-74\pm11$ \\
Dates observed & 2013 Apr.\ 12, 13 & 2013 Aug.\ 31; Sep.\ 3, 4  \\
Number of exposures & 7 & 7  \\
Total exposure time (s) & 12\,600 & 12\,600  \\
Seeing FWHM (\arcsec) & 2.2 & 2.1  \\
\hline
\end{tabular}
\end{table*}

The raw data were reduced using the standard SAMI data reduction pipeline, described in \citet{Sharp15}. This pipeline carries out all steps from raw data to the production of fully calibrated datacubes. The software itself is available online through its listing in the Astrophysics Source Code Library \citep{Allen14}; changeset \texttt{6c0c801}, the same as for the SAMI Galaxy Survey Early Data Release \citep{Allen15}, was used to produce the data used in this paper. A full assessment of the quality of the pipeline's results is given in \citet{Allen15}.

The stellar kinematics of each galaxy were measured from the SAMI datacubes using pPXF \citep{Cappellari04,Cappellari12}, following the procedure described in \citet{Fogarty14}. Briefly, the 985 MILES stellar templates \citep{SanchezBlazquez06} were fit to a spectrum extracted from the centre of the galaxy, to form a single master template. This master template was then fit to each spaxel individually, giving measurements of the velocity, velocity dispersion and higher moments at each location in the galaxy. Full details are given in \citet{Fogarty14}.

In order to measure the emission-line properties from the SAMI spectra, the stellar continuum must first be subtracted. In principle this could be done using the fits derived during the stellar kinematics measurements, described above, but these are not optimised for accurate subtraction of stellar features in the spectra. To ensure that features such as the Balmer absorption lines are best removed, we re-fit the stellar continuum using a subset of 65 of the 300 available single stellar population (SSP) synthetic spectra from the MILES model library \citep{Vazdekis10}, using the full subset at each spaxel. These model spectra form a regular grid of log(age) from 0.063 to 17.8\,Gyrs and metallicity, [$Z$/H], from -1.31 to 0.22 (equivalent to $Z = 0.001$ to 0.03). The 65 templates cover 13 different ages at each of 5 metallicities; tests with the full set of 50 ages showed no significant difference in the continuum fits. We include a multiplicative polynomial in the fit, in order to account for dust extinction and uncertainties in the flux calibration. Further details will be presented in Schaefer et al.\ (in prep.).

Following subtraction of the stellar continuum, the emission lines were measured using LZIFU, an IDL package developed for the SAMI Galaxy Survey \citep{Ho14}. LZIFU fits emission lines as the sum of one or more kinematic components, using the MPFIT implementation of the Levenberg-Marquardt least-square algorithm \citep{Markwardt09}. Each component consists of a Gaussian emission line for each of the requested transitions, with all transitions given the same velocity and velocity dispersion. Certain known flux ratios are also held fixed. To account for an uncertainty in the wavelength calibration of the blue arm of the spectrograph \citep{Allen15}, an additional velocity offset is applied to all lines in that arm, with the magnitude of the offset left as a free parameter in the fit. The result is a map of velocity and velocity dispersion for each kinematic component, along with maps of the flux for each emission line in each component. The emission-line fluxes were corrected for Milky Way dust extinction using the \citet{Schlafly11} correction to the \citet{Schlegel98} $E(B-V)$ maps, and a \citet{Fitzpatrick99} extinction curve.

For the centre of each galaxy we increased the signal-to-noise ratio (S/N) of the emission-line measurements by calculating the mean spectrum within a radius of 1\arcsec, taking into account the correlations between nearby spaxels [see \citep{Sharp15} for a discussion of this issue]. The stellar continuum and emission lines were then fit again using the same procedure.

We derived mass-weighted non-parametric star formation histories for our two objects using a spectral fitting technique following \citet{McDermid15}. Our approach will be described in detail in Scott et al.\ (in prep.), but we summarise it here. 

High S/N spectra were created by binning all spaxels within linearly-spaced elliptical annuli from the blue SAMI cubes for each object. Using pPXF, we fit to each spectrum 296 MILES SSP synthetic spectra with age from 0.0708 to 17.8\,Gyr and [$Z$/H] from -1.71 to 0.22, again including a multiplicative polynomial. For each fit we imposed a smoothness constraint (using the \texttt{REGUL} keyword in pPXF) to the weights of neighbouring models. The relative weights assigned to each SSP model give the relative mass of each age and metallicity combination; we converted these to absolute masses by scaling each spaxel's weights by the total stellar mass of the galaxy multiplied by the fraction of the galaxy's continuum light (rest-frame 6700\,\AA) in that spaxel. For the current work we summed over all metallicity values. The measurements presented here used only the blue arm of the SAMI spectra in the fit, but performing the same measurements on the combined blue and red spectra produced no significant difference in the results.

Radio observations of the galaxies were obtained from the FIRST survey \citep{Becker95}. We searched for radio sources within 30\arcsec{} of the coordinates of each galaxy, finding a match for 272831 but not for 9016800393. For 272831 we confirmed the match by a visual inspection of an overlaid image.

\begin{figure*}
\centering
\includegraphics[width=170mm]{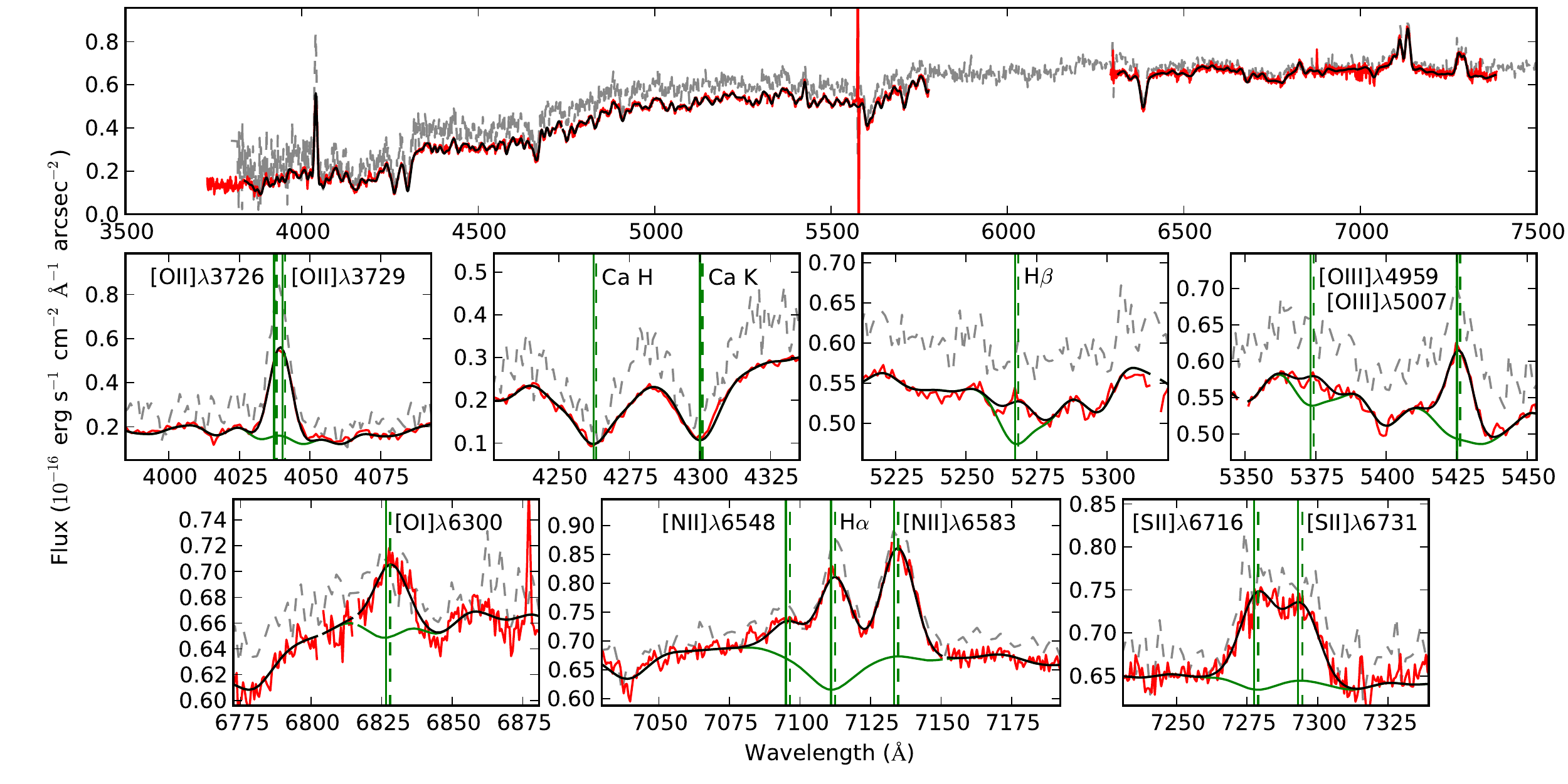} 
\caption{Observed spectrum (red) and one-component LZIFU fit (black) for a central pixel from 272831. In the lower panels, seven wavelength regions around key emission and absorption lines are shown. These panels include the pPXF continuum fit (as used for the LZIFU emission-line fit) in green, and the wavelengths of key lines at the OSSY host galaxy (solid vertical lines) and emission line (dashed vertical lines) redshifts, as labelled on the plot. Note that the $y$ range is different in each panel, and does not begin at zero. In all panels the grey dashed line shows the SDSS spectrum, scaled by the fibre area to place it on the same flux scale.}
\label{fig:spectrum_272831}
\end{figure*}

Both galaxies have FUV and NUV magnitudes available from the GALEX surveys \citep{Martin05,Morrissey07}. As 272831 lies within the footprint of the Galaxy And Mass Assembly (GAMA; \citealt{Driver09,Driver11}) survey we obtained observed magnitudes and k-corrections from the GAMA website\footnote{http://www.gama-survey.org}, and corrected them for Milky Way dust extinction using the same procedure as for the emission lines. For 9016800393 we corrected for Milky Way dust and then calculated rest-frame magnitudes using {\sc kcorrect} version v4\_2 \citep{Blanton07}.

Neither galaxy is found in the ROSAT All-Sky Survey catalogues \citep{Voges99}, and other X-ray observations are not available.

\section{Results}

As we will show in the following results, the observed properties and physical interpretations are very different for each of the two galaxies. Hence, we describe each one individually in the following subsections. 

\subsection{272831}

The mean spectrum from the centre of 272831 is shown in Fig.~\ref{fig:spectrum_272831}. The lower panels show seven key wavelength regions in more detail, including both the observed spectrum and the pPXF/LZIFU fit. We use a single kinematic component to describe the observed emission lines, as a second component does not provide significant detections (S/N$>$3) of all of the strong lines. Vertical lines in the figure mark the positions of prominent emission and absorption features at the redshifts given in the OSSY catalogue.

Fig.~\ref{fig:spectrum_272831} also shows the SDSS spectrum for 272831. There is some disagreement between the overall shape of the SDSS and binned SAMI spectra, but this is expected given that the spectra are formed from different apertures. The emission-line properties show a very close match between the two spectra.

\subsubsection{Velocity fields}

\begin{figure*}
\centering
\includegraphics[width=170mm]{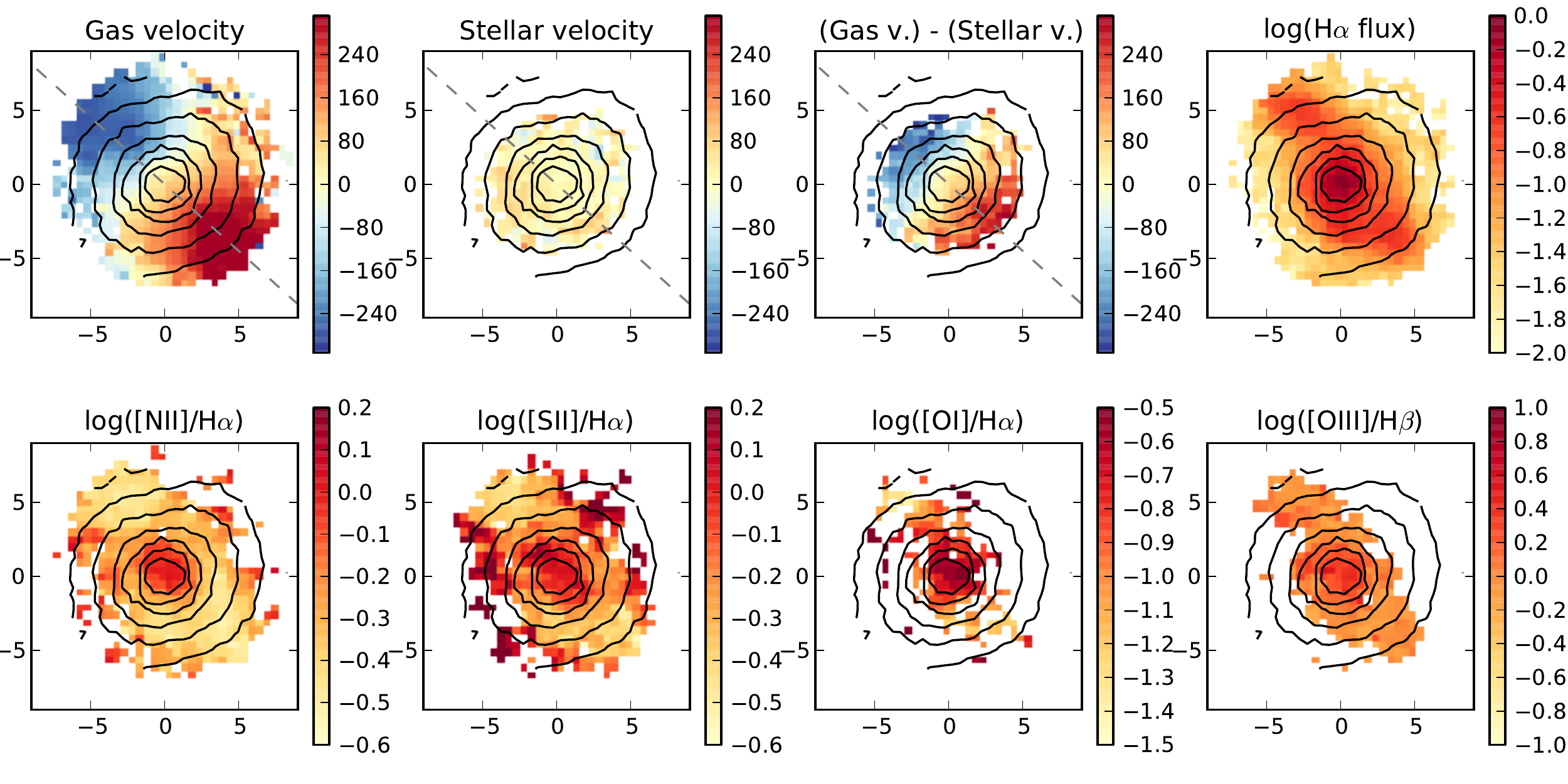} 
\caption{Kinematic and emission-line flux maps for the galaxy 272831. From left to right, the top panels show: ionised gas velocity field; stellar velocity field; difference in velocities between gas and stars; log(H$\alpha$ flux). The bottom panels show: log(\nii/H$\alpha$); log(\sii/H$\alpha$); log(\oi/H$\alpha$); log(\oiii/H$\beta$). All velocities are in units of \kms, while the H$\alpha$ flux is in units of $10^{-16}$\,erg\,s$^{-1}$\,cm$^{-2}$\,pix$^{-1}$. Contours show the continuum flux at rest-frame 6400\,\AA, as measured from the SAMI datacube, at intervals of 0.25\,dex. In each of the first three panels, a dashed line shows the kinematic axis of the gas. In all panels the spatial coordinates are in arcseconds, with North up and East to the left.}
\label{fig:kinematics_272831}
\end{figure*}

The emission-line and stellar velocity fields for 272831 are shown in Fig.~\ref{fig:kinematics_272831}, along with the difference between the two. The gas velocity is taken from the LZIFU measurements, which simultaneously fit all strong emission lines with a single velocity. It is immediately apparent that the ionised gas and stars have very different kinematic properties: the stellar velocity field has very little rotation, but there is strong ordered rotation in the ionised gas.

From the stellar properties we measure $\lambda_{\rm R}$, a parametrization of the specific angular momentum measured within one effective radius \citep{Cappellari07}, as $0.100\pm0.014$. Comparing to the \citet{Emsellem11} analysis of the kinematics of early-type galaxies in the ATLAS$^{\rm 3D}$ sample, 272831 is a slow rotator as measured in the stars, lying below the cutoff of 0.121 for its observed ellipticity of 0.153. The values of $\lambda_{\rm R}$ and ellipticity lie comfortably within the range observed by \citet{Emsellem11} for slow rotators, which make up 14 per cent of all early-type galaxies. The ATLAS$^{\rm 3D}$ sample shows a higher slow-rotator fraction for high-mass (above $10^{11}$\,M$_\odot$) galaxies like 272831; hence the stellar kinematic properties of 272831, while uncommon in the context of the galaxy population as a whole, are not exceptional.

The weak rotation of the stars is misaligned relative to the gas, with a difference in position angles of $66\pm3$\textdegree. The velocity field of the gas appears smooth, although we are not sensitive to structure on scales smaller than the seeing (FWHM 2\farcs2).

\begin{figure}
\centering
\includegraphics[width=85mm]{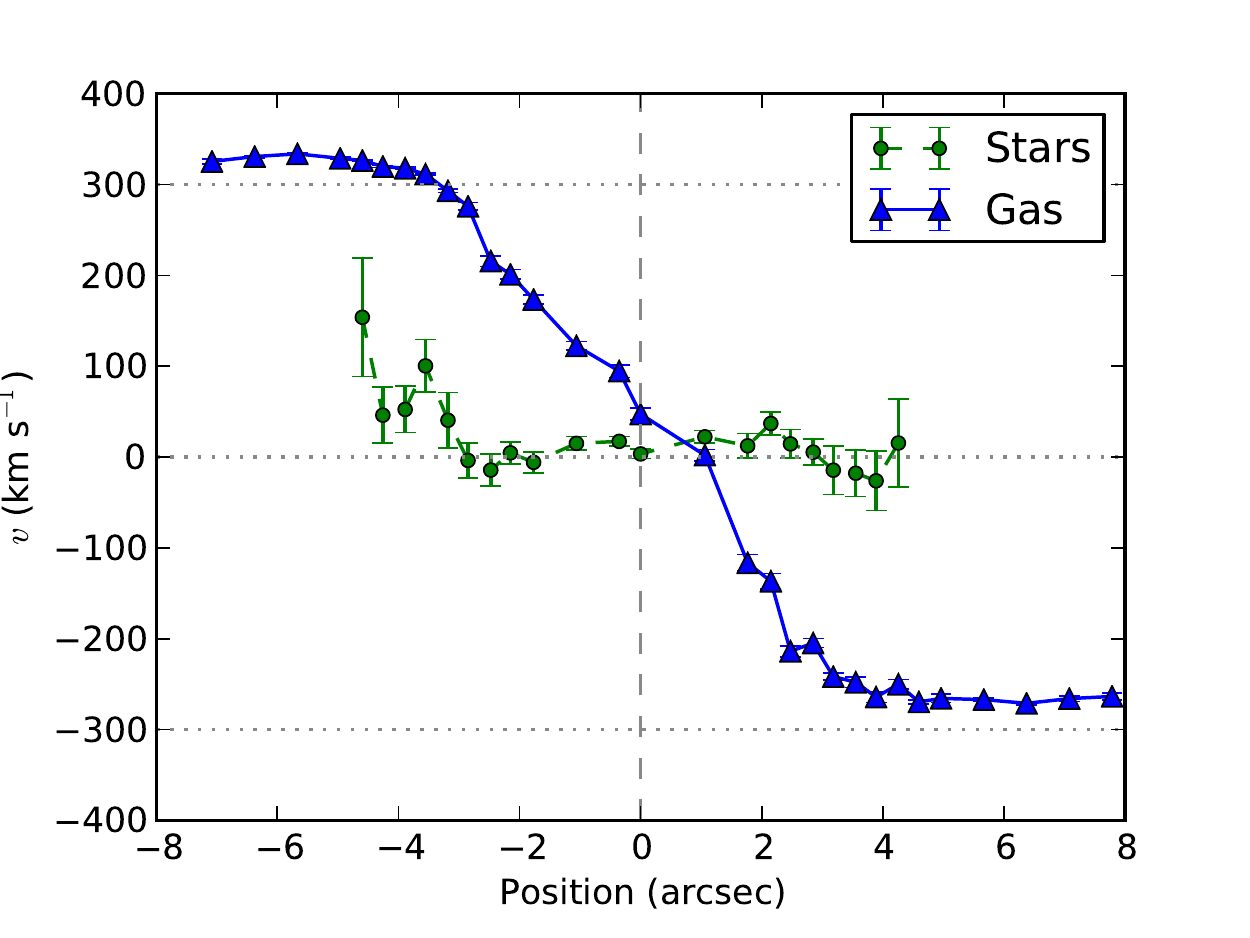} 
\caption{Velocity as a function of position in the galaxy for 272831. The two components plotted are stars (green circles, dashed line) and ionised gas (blue triangles, solid line). The vertical dashed line shows the centre of the galaxy. The cut is taken along the kinematic major axis of the ionised gas, with the positive direction lying approximately South-West. Dotted horizontal lines show the systemic velocity and, for illustration, $\pm300$\,\kms.}
\label{fig:velocity_profile_272831}
\end{figure}

Fig.~\ref{fig:velocity_profile_272831} shows the velocity as a function of position in the galaxy, taking a cut along the kinematic major axis of the ionised gas, approximately NE--SW. It is again clear that the gas and stars have very different kinematics -- strong rotation in the ionised gas but not in the stars -- but a further feature can now be seen. In the photometric centre of the galaxy, the ionised gas does not coincide in velocity with the stars, but is redshifted by $61\pm8$\,\kms, in agreement with the OSSY measured offset of $67\pm13$\,\kms. Similarly, the velocities of the two flat regions in the ionised gas velocity curve, at $+325\pm3$\,\kms{} and $-265\pm4$\,\kms, are not symmetric around zero, but have a mean offset of $30\pm3$\,\kms.

\subsubsection{Ionised gas properties}

\label{sec:ionisation_272831}

\begin{figure*}
\centering
\includegraphics[width=170mm]{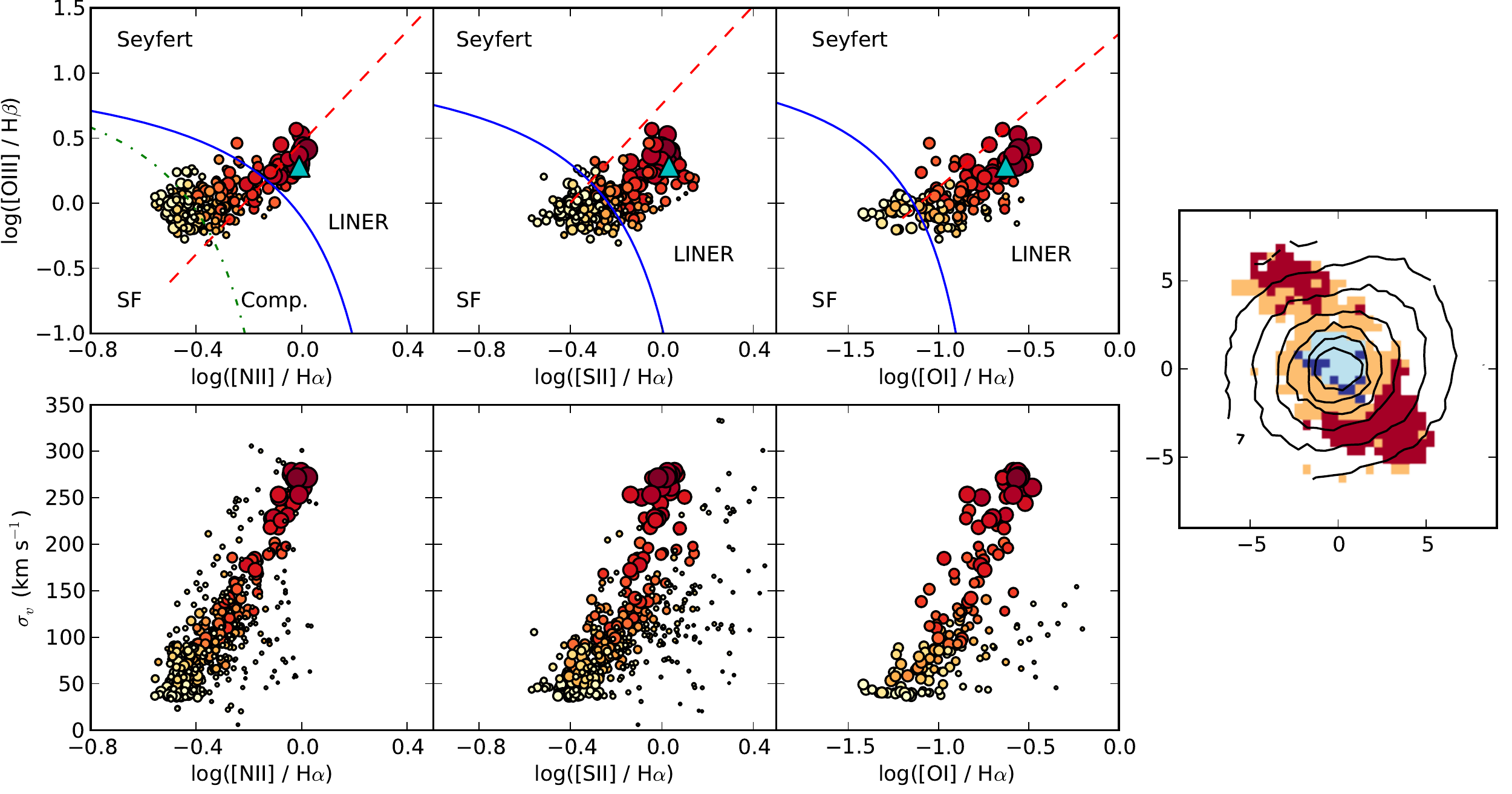} 
\caption{Ionised gas properties for 272831. Top row: line ratio diagrams for \nii{}, \sii{} and \oi{}, from left to right respectively. Bottom row: line width as a function of \nii/H$\alpha$, \sii/H$\alpha$ and \oi/H$\alpha$, from left to right respectively. In each panel the size of the points is proportional to their H$\alpha$ flux, and the colour shows the distance from the centre of the galaxy (red at centre, yellow at edge). In each panel, only points with S/N$>$3 in each of each of the lines featured in that panel are included. Cyan triangles on the line ratio diagrams show the OSSY measurements of the SDSS spectrum. The blue solid lines show the theoretical maximum starburst from \citet{Kewley01}, the green dot-dashed line shows the empirical starburst boundary from \citet{Kauffmann03}, and the red dashed lines show the Seyfert/LINER boundaries from \citet[\nii{} plot]{Kauffmann03} and \citet[\sii{} and \oi{} plots]{Kewley06}. The \citet{Kewley06} classification for each region is labelled. The right-hand panel shows the mapping of \citet{Kewley06} classifications back to the image of the galaxy: each spaxel is coloured according to its \nii{} classification as Seyfert (dark blue), LINER (light blue), composite (light orange) or star forming (dark red). Contours are as in Fig.~\ref{fig:kinematics_272831}.}
\label{fig:bpt_272831}
\end{figure*}

Further panels in Fig.~\ref{fig:kinematics_272831} show the distribution of H$\alpha$ flux and key diagnostic line ratios across 272831. Fig.~\ref{fig:bpt_272831} shows the same data displayed in a series of ionisation diagnostic diagrams (IDDs), including also the velocity dispersion of the gas \citep{Baldwin81,Veilleux87,Rich11}. The central spaxels lie in the `LINER' region of the \oiii/H$\beta$ vs.\ \nii/H$\alpha$, \oiii/H$\beta$ vs.\ \sii/H$\alpha$ and \oiii/H$\beta$ vs.\ \oi/H$\alpha$ diagrams, suggesting that the central gas may be ionised by shock heating \citep{Kauffmann03,Kewley06}. Photoionisation by an AGN accretion disc, or shock heating in the immediate vicinity of such a disc, are disfavoured due to the spatial extent of the ionised gas: fitting a two-dimensional Gaussian to the H$\alpha$ flux in the spaxels that lie above the \citet{Kewley01} \nii/H$\alpha$ classification line gives a FWHM of 3\farcs3, significantly larger than the 2\farcs2 seeing. Similar properties can also be caused by photoionisation by old stars \citep{Binette94,Singh13}, but this is unlikely to be the case in 272831 as the emission-line flux has a very different distribution to that of the stars. In the outer regions of the galaxy the location on the IDDs is consistent with photoionisation by star-forming \hii{} regions.

Given the weakness of the H$\beta$ emission line relative to the continuum level, it may be strongly affected by uncertainties in the template fit to the continuum. Such uncertainties would carry through to the \oiii/H$\beta$ ratio and hence the IDD classification. However, we can place strong limits on the allowed range of H$\beta$ fluxes based on the strength of H$\beta$ absorption observed in passive galaxies and the expected Balmer decrement (the H$\alpha$/H$\beta$ ratio).

The equivalent width (EW) in absorption of the H$\beta$ line in the template fit to the central spectrum is 1.82\,\AA, while the value measured of the observed spectrum (continuum and emission line) is 0.58\,\AA. If the template fit is inaccurate and the true EW is lower, the H$\beta$ emission line flux will in turn be lower. However, the fitted value is already towards the lower end of the range observed in passive galaxies. At the most extreme, we can take the value of $\simeq$1.25\,\AA{} observed in NGC4486 \citep{Vazdekis10}, which would reduce the emission-line flux by a factor of $(1.82 - 1.25) / (1.82 - 0.58) = 0.46$, or an increase of 0.34 in log(\oiii/H$\beta$). Such a change would move the central points in Fig.~\ref{fig:bpt_272831} into the Seyfert region of the \oiii/H$\beta$ vs.\ \nii/H$\alpha$ diagram (top left panel), but would not change the classification in the other diagrams.

An upper limit to the H$\beta$ flux comes from the expected Balmer decrement of 2.86 \citep{Osterbrock06}, for the extreme case of a galaxy with no dust reddening. Comparing to the observed value of 4.75, and noting that the H$\alpha$ flux is much less sensitive to incorrect stellar templates, the true H$\beta$ flux cannot be more than $4.75 / 2.86 = 1.66$ times greater than the measured value. Taking this limit would reduce the measured log(\oiii/H$\beta$) by 0.22, in which case the central spaxels would still lie within the LINER regions in Fig.~\ref{fig:bpt_272831}. In summary, despite the weakness of the H$\beta$ emission line we are able to place strong constraints on the line ratios, giving a confident classification of the galaxy.

All the points in the three IDDs lie close to the fiducial shock line defined by \citet{Sharp10}, which traces the properties of a galaxy, NGC 1482, showing ionisation by shock heating. In NGC 1482 the shocks are associated with a starburst-driven outflow. There is an important difference in that the shock galaxies in \citet{Sharp10} showed increasing \nii/H$\alpha$ with increasing radius, while 272831 shows decreasing \nii/H$\alpha$ with increasing radius, i.e.\ the spatial distribution is inverted. This difference reflects the different sources of heating in the two galaxies: unlike 272831, NGC 1482 shows no signature of an AGN, indicating that the outflow is propelled and heated by star formation. Similar line ratios were also seen in a galaxy observed as part of the SAMI commissioning \citep{Fogarty12}, and interpreted as a shock heating due to an outflow driven by star formation; again the spatial distribution of ionisation properties was reversed relative to 272831.

A further galaxy from the SAMI Galaxy Survey with related ionisation properties is 209807 \citep{Ho14}, again interpreted as star-formation-driven shock heating. This interpretation was reinforced by a strong correlation between the ionised gas velocity dispersion and the \nii/H$\alpha$, \sii/H$\alpha$ and \oi/H$\alpha$ ratios, which is also seen in 272831, as shown in the fourth, fifth and sixth panels of Fig.~\ref{fig:bpt_272831}. In 209807 the higher ionisation gas was detected in the centre of the galaxy, but only in the form of a distinct kinematic component forming the outflow. We do not detect a separate component of this sort in 272831; a single component is sufficient to describe the emission across the entire galaxy. In summary, although the IDDs are consistent with shock-heated gas, the kinematic information suggests a different configuration to the outflows seen in other similar galaxies.

We measure the gas-phase metallicity of 272831 using the combined \nii/H$\alpha$ and $R_{23} = ($\oii\,$\lambda$3727 + \oiii\,$\lambda\lambda$4959,5007)/H$\beta$ calibration of \citet{Mannucci10}, in turn based on the calibrations of \citet{Maiolino08}. We choose this calibration in order to allow direct comparison with the \citet{Mannucci10} mass--metallicity relation. In the outer regions of 272831, where the emission-line properties are consistent with star formation, the metallicity is $12+\log({\rm O/H})=8.93\pm0.04$, with no obvious spatial variation. Using each of the \nii/H$\alpha$ and $R_{23}$ calibrators in isolation gives similar values: $8.87\pm0.03$ and $8.98\pm0.02$, respectively.

We estimate the Eddington ratio of the AGN from the observed \oiii{} luminosity, using the \oiii-to-bolometric conversion $L_{\rm bol}=3500\,L_{\rm \oiii}$ from \citet{Heckman04}. Extracting the \oiii{} luminosity from a circular aperture with radius 2\arcsec{} centred on 272831, we find $L_{\rm bol}=1.7\times10^{10}\,{\rm L}_\odot$, with an uncertainty of 0.38\,dex due to the scatter in the $L_{\rm bol}$--$L_{\rm \oiii}$ relation. Note that this luminosity, and hence the derived Eddington ratio, is likely an overestimate if some or all of the central \oiii{} flux arises from shocks or other ionisation sources. We measure a central stellar velocity dispersion, $\sigma_v$, of $249\pm6$\,\kms, and use the \citet{Gultekin09} relation to convert this to a black hole mass, $M_{\rm BH}$, of $4.0\times10^8$\,M$_\odot$. The scatter in the $M_{\rm BH}$--$\sigma_v$ relation gives an uncertainty of 0.31\,dex in $M_{\rm BH}$. Combining the $L_{\rm bol}$ and $M_{\rm BH}$ measurements, we derive an Eddington ratio of $1.3\times10^{-3}$, with an uncertainty of 0.49\,dex.

\subsubsection{Stellar populations}

\label{sec:stellar_pops_272831}

\begin{figure}
\centering
\includegraphics[width=85mm]{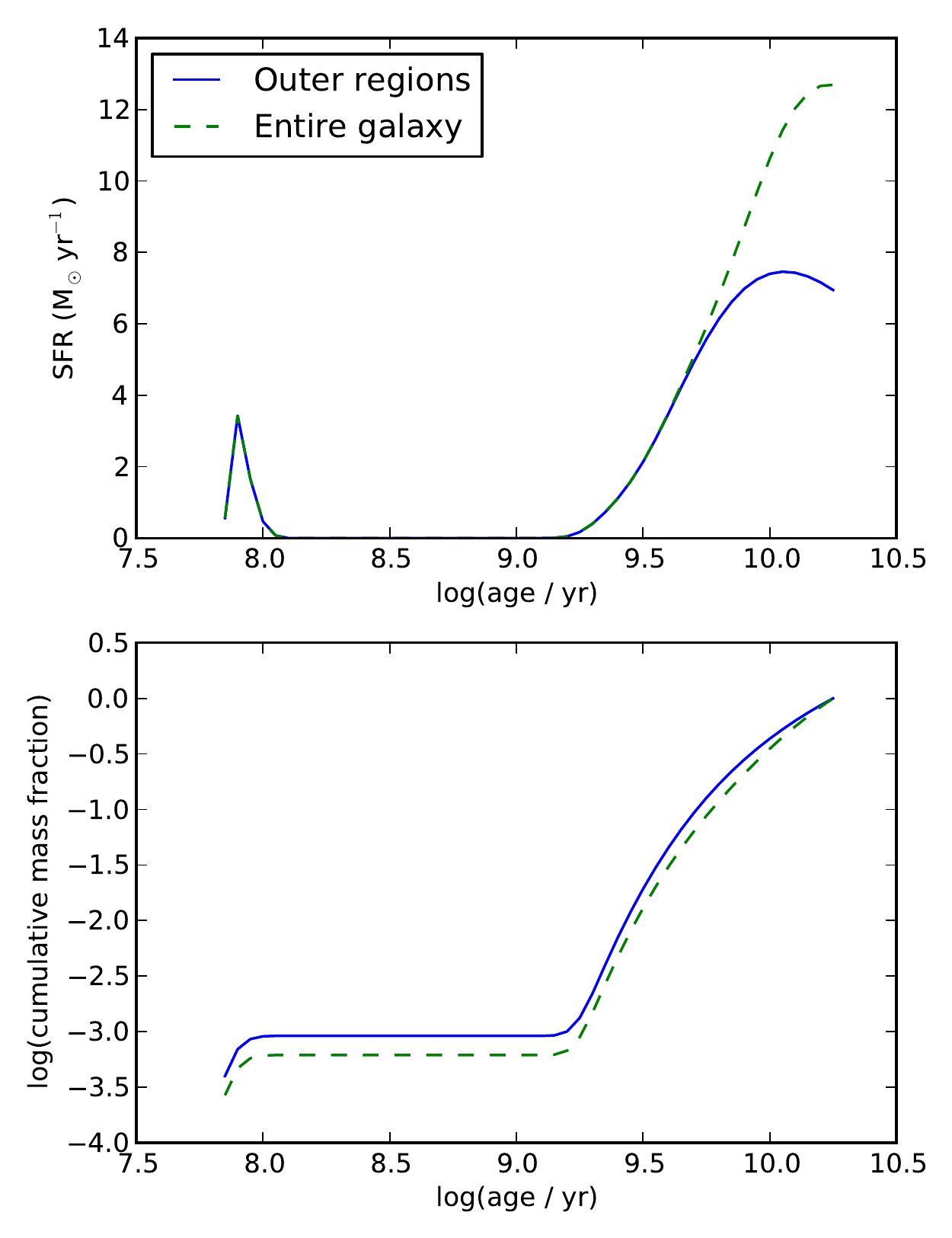} 
\caption{Upper panel: SFR as a function of age, as measured from the stellar populations, for the entirety of 272831 (solid blue line) and only the outer regions (dashed green line). Lower panel: Cumulative mass fraction as a function of age, i.e. the fraction of stellar mass in stars younger than the $x$-axis value.}
\label{fig:stellar_pops_272831}
\end{figure}

The upper panel of Fig.~\ref{fig:stellar_pops_272831} shows the star formation rate (SFR) as a function of age for 272831, measured from the stellar populations fits. The lower panel shows the cumulative stellar mass as a fraction of the total, plotted as a function of age. The vast majority of the stellar mass was formed more than 1.6\,Gyr ($10^{9.2}$\,yr) ago, with a small burst of star formation occurring within the last 0.1\,Gyr. This recent burst has formed less than 0.1 per cent of the total mass of the galaxy.

In order to compare the stellar populations results to the current SFR measured from H$\alpha$ we discard the central region, up to a radius of 1\farcs7, and retain only the outer regions of the galaxy. Doing so avoids contamination by the AGN or shocks. The discarded region is 1.5 times the size of the seeing disc for the data, and corresponds to a physical size of 2.7\,kpc. It encloses the region of high \nii/H$\alpha$ seen in Fig.~\ref{fig:kinematics_272831} so can be assumed to contain the majority of emission from gas ionised by sources other than star formation. The remaining emission in the retained region is dominated by star formation. After correcting for dust extinction using the observed Balmer decrement and the $\mu_{0.4}$ dust curve from \citet{Wild11}, we sum the extinction-corrected H$\alpha$ flux in this region. We then convert this flux to an SFR using the relation in \citet{Kennicutt98}, converted to a \citet{Chabrier03} initial mass function (IMF), giving $0.66\pm0.04$\,M$_\odot$\,yr$^{-1}$. If the H$\alpha$ flux across the entire field of view is assumed to be due to star formation, the SFR increases to $1.05\pm0.05$\,M$_\odot$\,yr$^{-1}$, but this is certainly an overestimate due to contamination by other ionisation sources in the centre of the galaxy.

The H$\alpha$ SFR is close to the SFR inferred from the youngest stars in the stellar populations fit: a mass of $4.0\times10^7$\,M$_\odot$ was found for the population with age 71\,Myr, corresponding to an SFR of 0.56\,M$_\odot$\,yr$^{-1}$ averaged over the age of the bin. In this case no young stellar populations were found in the discarded central region of the galaxy, so the SFR is the same for the outer regions as for the entire galaxy.

\begin{figure*}
\centering
\includegraphics[width=170mm]{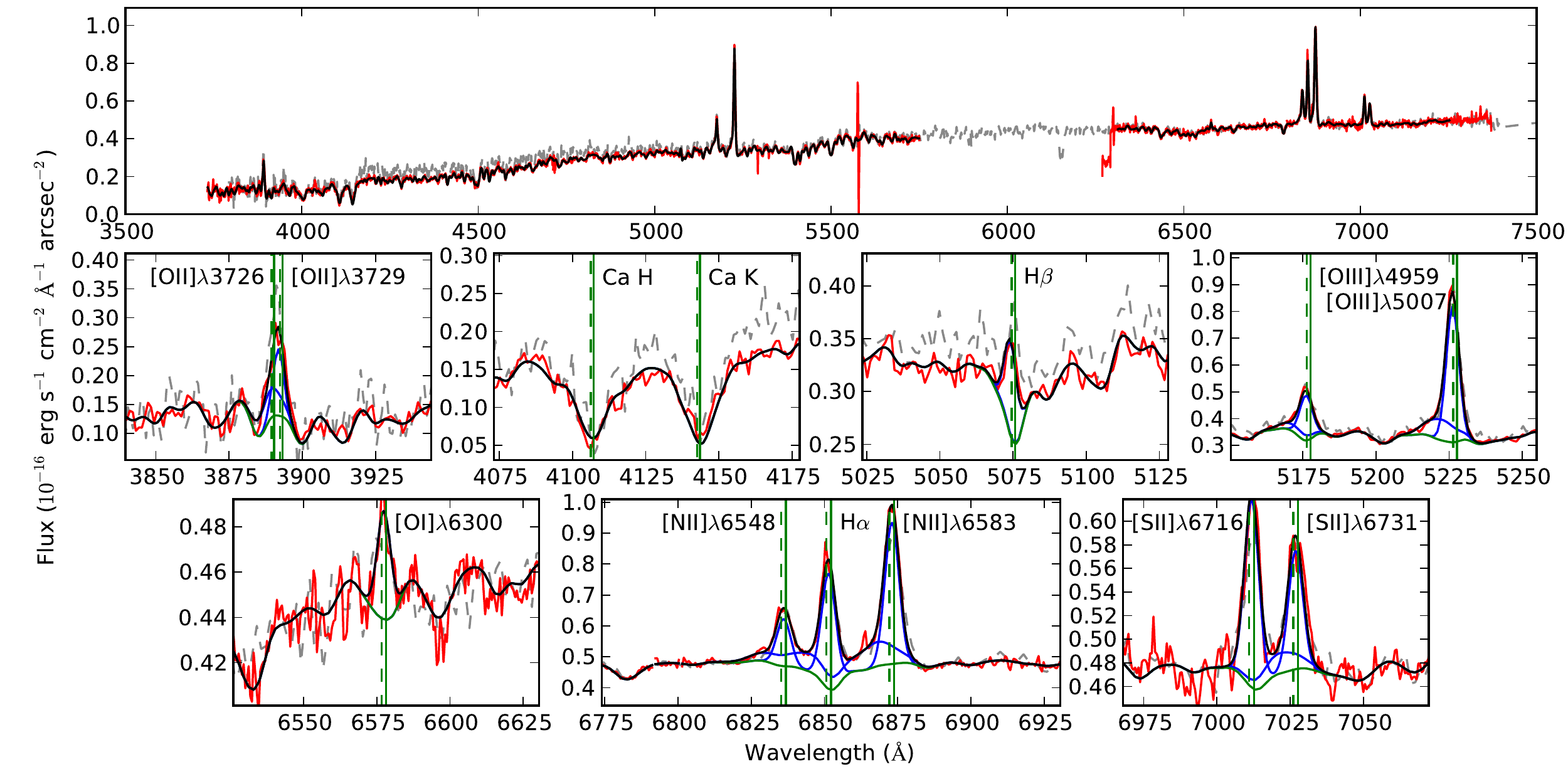} 
\caption{As Fig.~\ref{fig:spectrum_272831} but for the galaxy 9016800393. In this case a two-component LZIFU fit is used; each emission-line component, added individually to the stellar component, is plotted in blue. The lower panels include the LZIFU fits to each component.}
\label{fig:spectrum_9016800393}
\end{figure*}

\subsubsection{Radio emission}

The galaxy is detected in the FIRST radio catalogue, with an integrated flux at 1.4\,GHz of 4.27\,mJy, which corresponds to a radio luminosity of $7.54\times10^{22}$\,W\,Hz$^{-1}$. Using the calibration of \citet{Hopkins03} and assuming that all the radio continuum emission is due to star formation, this would correspond to an SFR of 41.7\,M$_\odot$\,yr$^{-1}$, far higher than the upper limit of 1.05\,M$_\odot$\,yr$^{-1}$ implied by the H$\alpha$ emission. The vast majority of the radio continuum must therefore be due to an AGN. The emission lines in the centre of the galaxy have an excitation index, as defined by \citet{Buttiglione10}, of $0.55\pm0.02$, below the cutoff of 0.95, making 272831 a low excitation radio galaxy (LERG). This classification is consistent with the radio luminosity, which is in the range dominated by LERGs \citep{Best12}, and with the low Eddington ratio of the AGN.

\subsubsection{UV emission}

The GALEX observed magnitudes for 272831 are $20.51\pm0.09$ and $19.65\pm0.04$ for the FUV and NUV filters, respectively. After removing the effect of dust extinction within the Milky Way, and k-correcting, we derived rest-frame magnitudes of $20.23\pm0.09$ and $19.12\pm0.04$. Using the calibration of \citet{Salim07} for `normal' galaxies, again assuming a \citet{Chabrier03} IMF, these correspond to an SFR of $12.8\pm3.0$\,M$_\odot$\,yr$^{-1}$. This calibration includes an empirical correction for intrinsic dust extinction, based on the FUV$-$NUV colour. The quoted uncertainty does not include the uncertainty in this correction, which increases the flux by a factor of 37, so will be an underestimate. However, the large dust correction is consistent with the relatively high Balmer decrement, $4.8\pm0.2$, seen in the centre of the galaxy.

The SFR calculated here may be contaminated by UV flux from the AGN, but \citet{Salim07} show that such contamination should not exceed 15 per cent of the total flux, based on an extrapolation of the limit on $r$-band contamination measured by \citet{Kauffmann03}. Conservatively assuming the maximum level of contamination gives a final UV SFR of $10.9\pm2.6$\,M$_\odot$\,yr$^{-1}$. This value is an order of magnitude higher than the H$\alpha$ SFR, which may be due to the different timescales probed by the two methods. The H$\alpha$ flux is measuring star formation more recent than the peak seen in Fig.~\ref{fig:stellar_pops_272831}, but the UV flux, which probes timescales up to a few 10$^8$\,yr, includes that peak and so is expected to give a higher SFR. The discrepancy between the two measures may also be enhanced by uncertainties in the dust correction, which can vary greatly between individual galaxies \citep{Boselli09}.

\subsection{9016800393}

\begin{figure*}
\centering
\includegraphics[width=170mm]{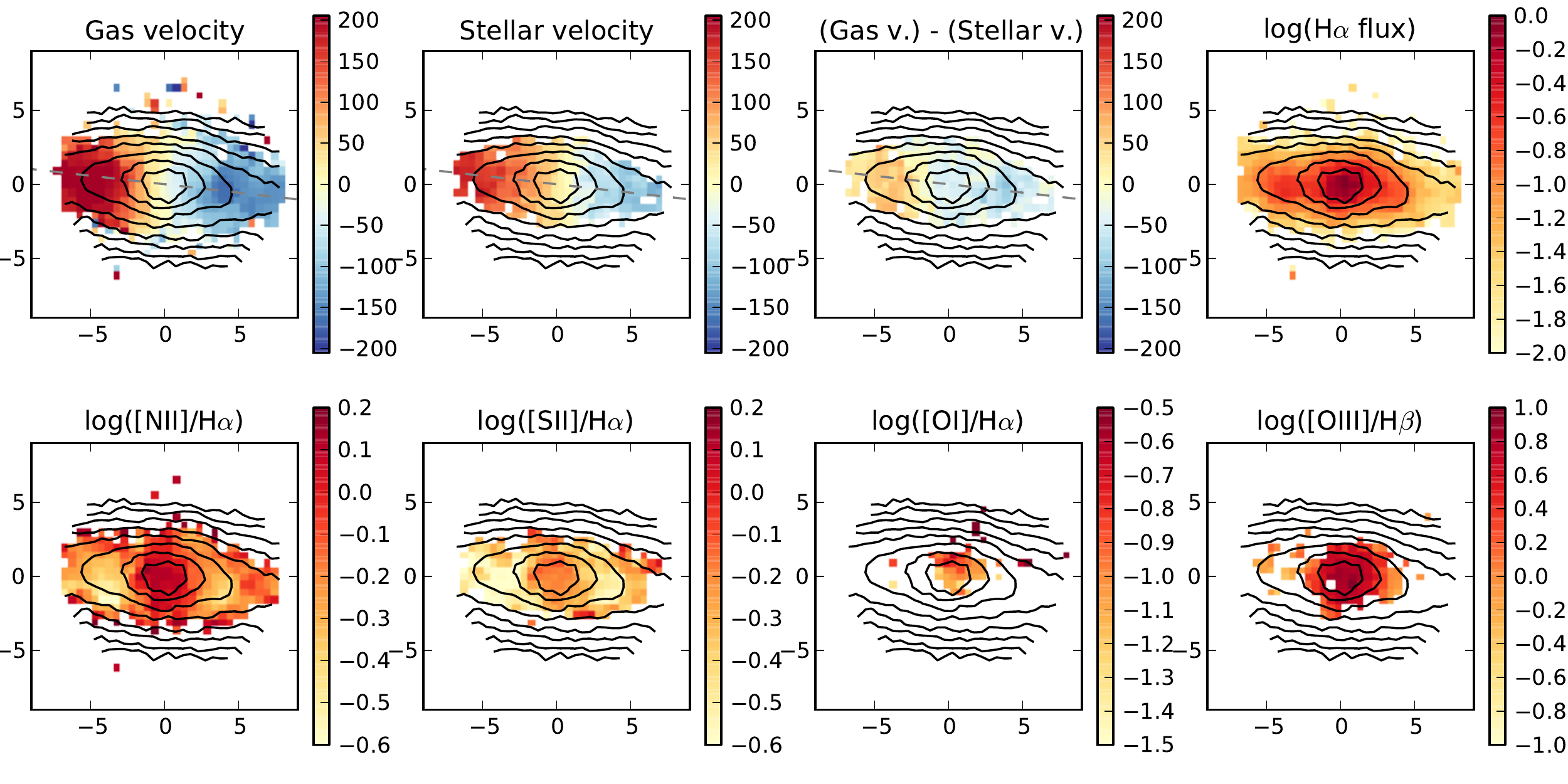} 
\caption{As Fig.~\ref{fig:kinematics_272831}, for the galaxy 9016800393. The measurements from the single-component LZIFU fit are shown.}
\label{fig:kinematics_9016800393}
\end{figure*}

The central spectrum of 9016800393 is shown in Fig.~\ref{fig:spectrum_9016800393}. In this case, two velocity components are required to provide a good fit to the emission lines: a narrow component \mbox{($\sigma_v=107\pm1$\,\kms)} with a blueshift of $44\pm2$\,\kms relative to the stars, and a broader component \mbox{($\sigma_v=304\pm7$\,\kms)} that is blueshifted by a further $229\pm14$\,\kms{} relative to the narrow component. The broader component is measured at high S/N, formally an 18-$\sigma$ detection in the \oiii\,$\lambda$5007 line. The blueshift measured by OSSY, $74\pm11$\,\kms, lies between those for the two components in the SAMI data, suggesting it was influenced by both physical components.

\subsubsection{Velocity fields}

The stellar and ionised gas velocity fields for 9016800393 are shown in the upper panels of Fig.~\ref{fig:kinematics_9016800393}. In the ionised gas we show for simplicity a single-component fit to the data. Even in the central region, the velocity measured in this way is dominated by the narrower of the two velocity components.

In contrast to 272831, both the ionised gas and the stars in 9016800393 show simple ordered rotation, with position angles that agree within the measurement uncertainties. In the outer regions of the galaxy (radius 5--7\arcsec), the velocity of the stars has a mean value of 79 per cent of that of the ionised gas, using the single-component LZIFU fits. This offset is as expected due to asymmetric drift, the tendency of stars to lag behind the gas in rotation velocity due to their increased random motions. The value observed in 9016800393 is close to the mean value of 75 per cent in the SAMI Galaxy Survey sample as a whole \citep{Cortese14}, and comparable to the $89\pm8$ per cent measured by \citet{Martinsson13} from 30 spiral galaxies in the DiskMass survey.

\begin{figure}
\centering
\includegraphics[width=85mm]{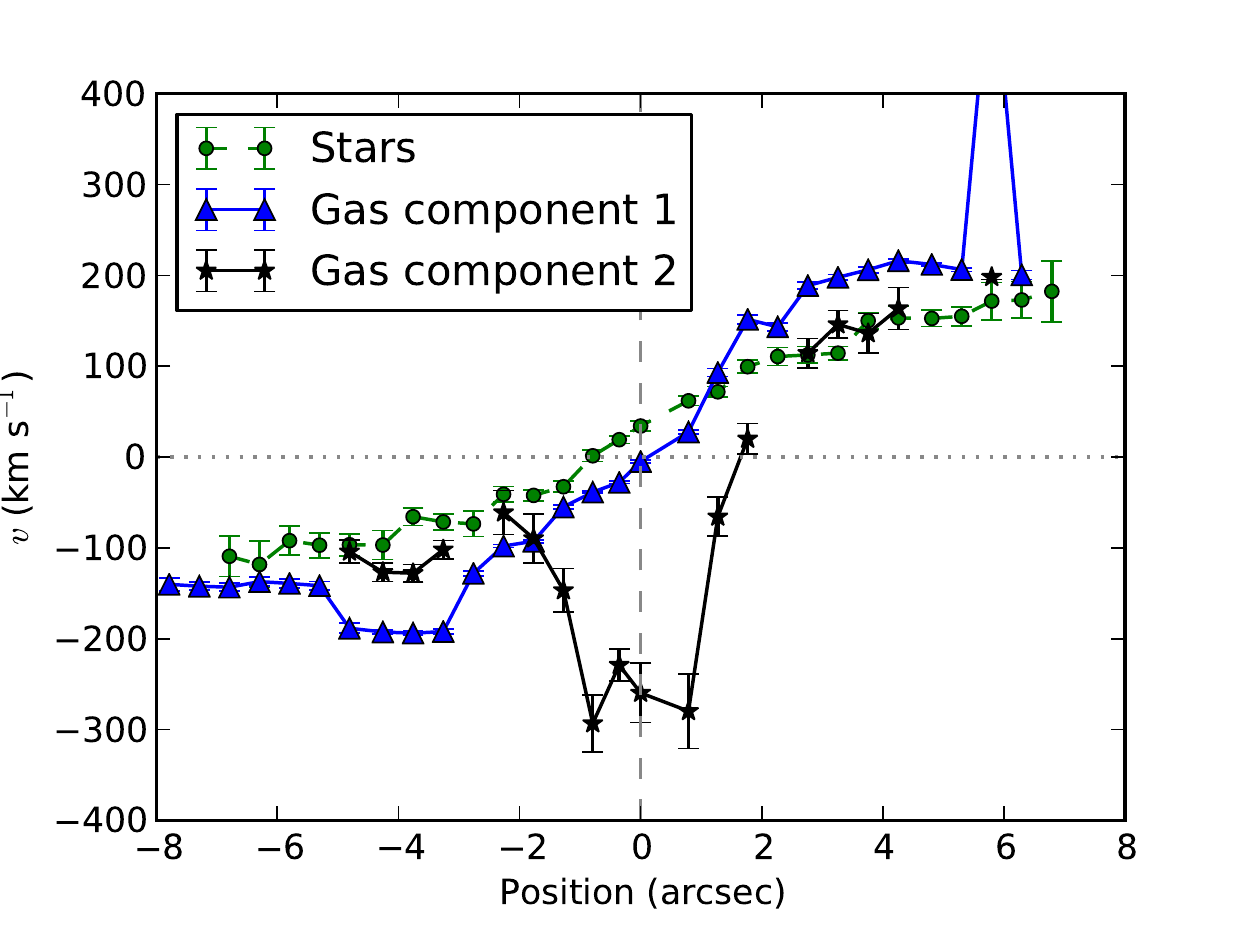} 
\caption{Velocity as a function of position in the galaxy for 9016800393. The components plotted are: stars (green circles, dashed line), narrow gas (blue triangles, solid line) and broad gas (black stars, solid line). The vertical dashed line shows the centre of the galaxy, while the horizontal dotted line marks the systemic velocity. The cut is taken along the kinematic major axis, with the positive direction lying approximately East.}
\label{fig:velocity_profile_9016800393}
\end{figure}

Fig.~\ref{fig:velocity_profile_9016800393} shows the velocity as a function of position in the galaxy for each of the components, taking a cut along the kinematic major axis. Each of the two components of the ionised gas is shown. These two components have markedly different behaviour: the narrow component traces the stellar kinematics closely, other than the asymmetric drift described above, but the broader component shows a blueshift of $\sim$250\,\kms{} in the centre of the galaxy, as seen in the single spectrum shown in Fig.~\ref{fig:spectrum_9016800393}.

A careful examination of the spectra in Fig.~\ref{fig:spectrum_9016800393} shows that the narrow component of the \oiii{} emission lines is blueshifted relative to the same component in other emission lines, with an offset of 71\,\kms, bringing its total blueshift relative to the stars to 115\,\kms. The LZIFU fitting procedure has incorporated this shift within the parameter that defines a wavelength calibration offset between the blue and red arms of the spectrograph. However, in this case the excellent agreement between the SAMI and SDSS emission line positions in both arms indicates that the offset is not due to a calibration error but a genuine shift in the \oiii{} velocity. The offset is not apparent in the other forbidden lines.

\subsubsection{Ionised gas properties}

\begin{figure*}
\centering
\includegraphics[width=170mm]{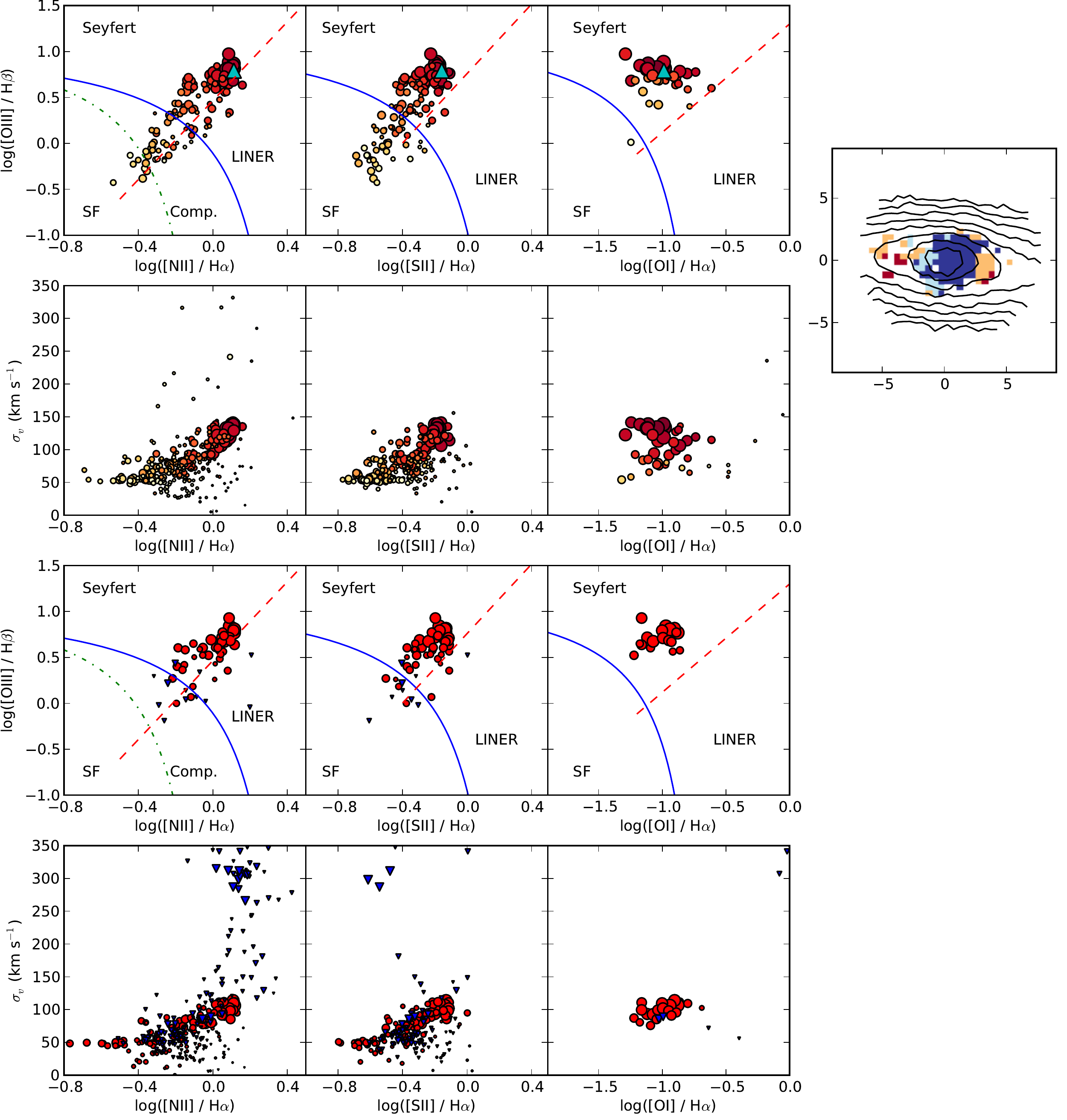} 
\caption{Ionised gas properties for 9016800393. The upper two rows are as in Fig.~\ref{fig:bpt_272831}, showing the IDDs for single-component fits to the emission lines. The panel on the right shows the mapping of \nii{}-based classifications back to the galaxy, using the same colour coding as in Fig.~\ref{fig:bpt_272831}. The lower two rows show the results for the two-component fits: the red circles show the narrow component and the blue triangles show the broad component. In each panel, each line in the diagram must have S/N$>$3 in the individual component for a point to be included.}
\label{fig:bpt_9016800393}
\end{figure*}

Maps of the emission-line properties in 9016800393 are shown in Fig.~\ref{fig:kinematics_9016800393}, using the single-component fits to the emission lines. The upper panels of Fig.~\ref{fig:bpt_9016800393} show the IDDs for the same fits. In the lower panels, the same diagrams are shown with the emission separated into the two kinematic components observed.

Looking first at the single-component fits, it is clear that the emission lines in the centre of 9016800393 are dominated by a Seyfert-like AGN, with some star formation in the outer disc. The points lie close to the fiducial AGN-excited line defined in \citet{Sharp10}. The 1-$\sigma$ width of the lines does not exceed 150\,\kms{} across the galaxy.

When two components are fit to the emission lines, as shown in the lower panels of Fig.~\ref{fig:bpt_9016800393}, the narrower component shows essentially the same results as for the single-component fits. The broad component in the outer spaxels also follows the single-component results, suggesting that in this region it may be due to seeing effects and integrating along the line-of-sight in an edge-on galaxy producing non-Gaussian emission line profiles, rather than being a separate physical component. However, in the central region the broader component shows some marked differences: it is significantly broader than any of the single- or narrow-component fits ($\sim$300\,\kms{} as opposed to $<$150\,\kms), the \oiii/H$\beta$ ratio is somewhat higher, and it has a lower \sii/H$\alpha$ ratio (logarithm $-0.6$, compared to $-0.2$ for the narrow component in the same region, lower by a factor of $\sim$4). The broader component is marginally spatially resolved: a 2D Gaussian fit to its spatial profile in \oiii{} gives a FWHM in the major/minor axes of 3\farcs4/2\farcs6, compared to a seeing FWHM of 2\farcs1 (1\arcsec{} corresponds to 876\,pc at the redshift of 9016800393). The extension of the broader component is aligned 11\textdegree{} anticlockwise of the E--W direction, or 18\textdegree{} away from the major axis of the galaxy.

The unusually low \sii/H$\alpha$ ratio for material excited by an AGN suggests the broad component may be collisionally de-excited, indicating that the electron density is at or above the critical density for the relevant transitions, $1.2\times10^3$ and $3.3\times10^3$\,cm$^{-3}$ for the 6716-\AA{} and 6731-\AA{} transitions, respectively \citep{Rubin89}. However, the limited S/N of the broad component measurement, which will be compounded by any uncertainties in the continuum subtraction, prevent us from placing strong constraints on the density.

The log(\nii/H$\alpha$) ratio in the outer regions of 9016800393 is $-0.29$, higher than the upper limit for reliable metallicity measurements in gas ionised by young stars \citep{Mannucci10}, so we use only the $R_{23}$ calibration. The resulting metallicity is $12+\log({\rm O/H})=8.97\pm0.07$.

Following the same procedure as in Section~\ref{sec:ionisation_272831} to derive the Eddington ratio for the AGN in 9016800393, we measure \mbox{$L_{\rm bol}=8.4\times10^9$\,L$_\odot$} and \mbox{$M_{\rm BH}=1.5\times10^7$\,M$_\odot$}. These combine to give an Eddington ratio of $1.7\times10^{-2}$, with an uncertainty of 0.58\,dex due to the scatter in the scaling relations.

\subsubsection{Stellar populations}

\begin{figure}
\centering
\includegraphics[width=85mm]{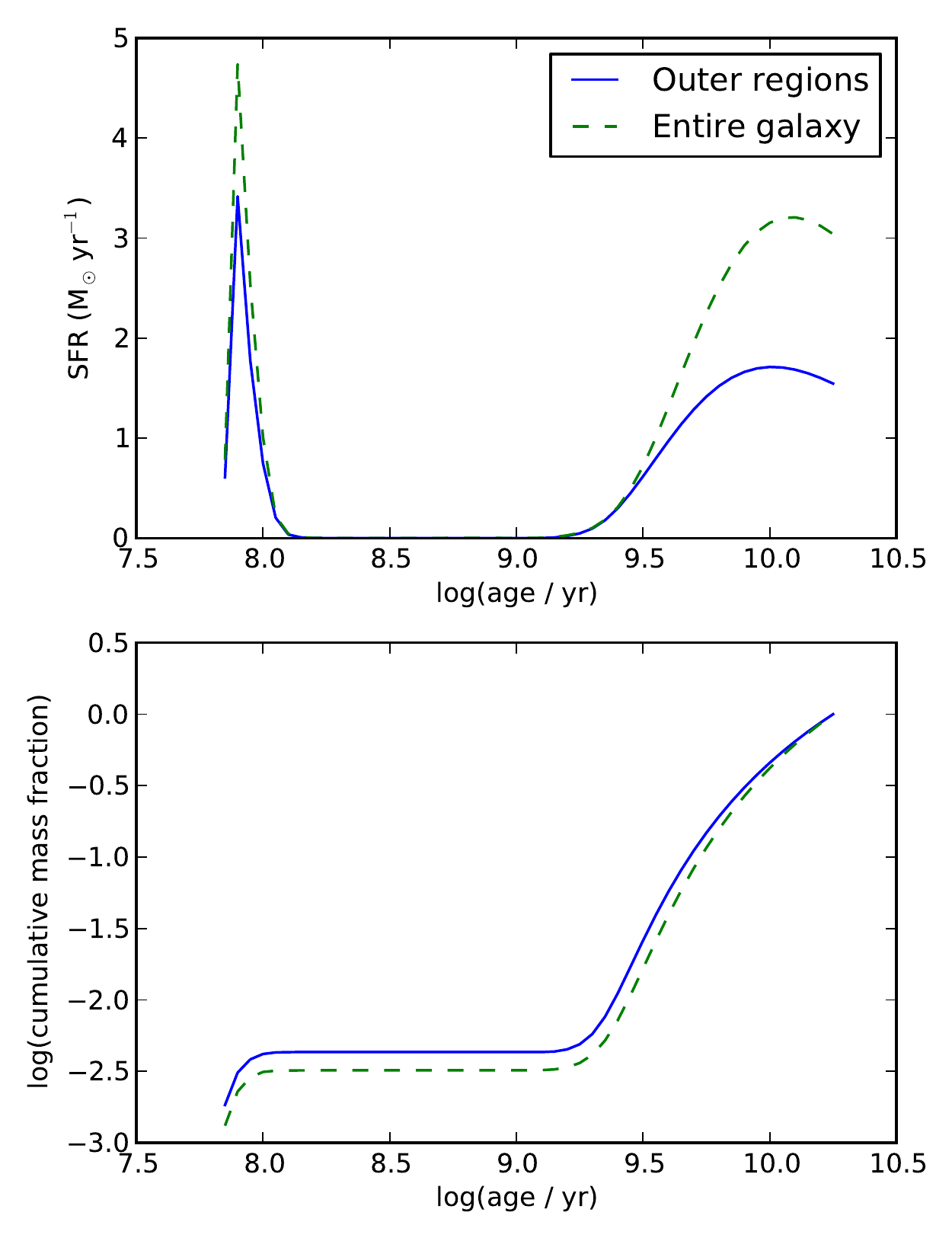} 
\caption{As Fig.~\ref{fig:stellar_pops_272831}, for the galaxy 9016800393.}
\label{fig:stellar_pops_9016800393}
\end{figure}

The SFR and cumulative mass fraction for 9016800393, from the stellar populations fits, are shown in Fig.~\ref{fig:stellar_pops_9016800393}. The shape of the SFR distribution as a function of time is similar to that of 272831, with an extended period of star formation at ages above 1.6\,Gyr, and a strong peak of recent star formation within the last 0.13\,Gyr. The recent star formation accounts for 0.3 per cent of the total mass of the galaxy.

Again discarding the central region of the galaxy, we find a mass in the youngest stars (71\,Myr) of $4.3\times10^7$\,M$_\odot$, corresponding to an averaged SFR of 0.61\,M$_\odot$\,yr$^{-1}$. The H$\alpha$ SFR for the same region is substantially lower, $0.060\pm0.007$\,M$_\odot$\,yr$^{-1}$. Integrating across the entire galaxy gives a mass of $5.5\times10^7$\,M$_\odot$, equivalent to an averaged SFR of 0.78\,M$_\odot$\,yr$^{-1}$, and an H$\alpha$ SFR of $0.263\pm0.019$\,M$_\odot$\,yr$^{-1}$. However, using the entire galaxy will strongly contaminate the H$\alpha$ SFR with the AGN emission.

\subsubsection{Radio emission}

There is no detection in FIRST for 9016800393, with a catalogue detection limit (including clean bias) at its coordinates of 0.86\,mJy, putting an upper limit on the radio power of the AGN at $4.0\times10^{21}$\,W\,Hz$^{-1}$. This limit is much higher than the radio power expected due to star formation, which lies in the range 1.1--$4.8\times10^{20}$\,W\,Hz$^{-1}$, depending on the fraction of H$\alpha$ emission arising from the AGN.

\subsubsection{UV emission}

The GALEX observed magnitudes for 9016800393 are $21.56\pm0.08$ and $20.80\pm0.07$ for the FUV and NUV filters, respectively. After removing the effect of dust extinction within the Milky Way, and k-correcting, we derived rest-frame magnitudes of $21.32\pm0.08$ and $20.48\pm0.07$. Using the calibration of \citet{Salim07} for `normal' galaxies, these correspond to an SFR of $0.54\pm0.15$\,M$_\odot$\,yr$^{-1}$. As before, the quoted uncertainty does not include the uncertainty in the intrinsic dust correction, so will be an underestimate. Correcting for contamination by the AGN gives a final UV SFR of $0.46\pm0.13$\,M$_\odot$\,yr$^{-1}$.

\section{Discussion}

\label{sec:discussion}

Again, due to the very different properties of the two galaxies revealed by the SAMI observations, we examine each one individually in the following subsections.

\subsection{272831}

The striking difference between the gas and stellar kinematics in 272831 implies an external origin for the gas, either from a recent merger with a gas-rich galaxy, or direct accretion. In either case the stellar populations indicate that the star formation associated with the event occurred within the last 100\,Myr. This age may not correspond directly to the timescale for the event itself as, for example, a merger may trigger star formation at first passage, as much as 1\,Gyr before the final coalescence \citep{Patton13}. The UV SFR is higher than either the H$\alpha$ SFR or the SFR inferred from the stellar populations, which may result from the uncertainty in the UV dust extinction correction, or simply the scatter in the relationship between UV luminosity and SFR.

Comparing to fig.~1 of \citet{Mannucci10}, the metallicity of the gas, $12+\log({\rm O/H})=8.93\pm0.04$, is somewhat low for a galaxy with a stellar mass of $10^{11.17}$\,M$_\odot$, lying at the edge of the 90 per cent range illustrated in the figure. The median metallicity at that mass is 9.16. The observed metallicity is more typical of a galaxy with a stellar mass in the range $10^{9.7}$--$10^{10.2}$\,M$_\odot$, possibly implying that the gas results from a merger with a galaxy of approximately this mass.

The kinematic centre of the gas is offset in velocity from the stars, resulting in the offset detected by CG14. This offset implies that the gas has yet to settle into a stable configuration, in agreement with the relatively short timescale for the accretion or merger event. The ionisation properties in the centre are consistent with shock heating, indicating that the ionisation may not be directly due to photoionisation from an AGN. Instead, it may be an indirect result, due to interaction of an AGN jet with the gas, or not related at all. In this last possibility, the gas may be shock heated due to interaction of the accreted/merged gas with any pre-existing gas in the galaxy. In either case, we do not know whether the AGN is moving with the velocity seen in the ionised gas, or is consistent with the stars.

We cannot rule out the possibility that 272831 contains two SMBHs. In particular, if the observed properties are due to an ongoing merger, if each of the merging galaxies contained an SMBH, and if those SMBHs have yet to coalesce, then the two SMBHs will currently be present. In this case, the SMBHs may be in a binary system or may not yet be under each other's direct gravitational influence. However, the assumptions that underpin this scenario cannot be verified with the available observations.

A galaxy with similar properties, SDSS J171544.0+600835 (hereafter J1715), was recently described by \citet{Villforth15}. In the case of J1715, double-peaked narrow emission lines were observed, suggesting a binary SMBH system in which both SMBHs are active, and this interpretation was previously confirmed by the detection of two distinct X-ray sources \citep{Comerford11}. The kinematics of J1715 appear very similar to those of 272831 (fig.~7 of \citealt{Villforth15}), in that the stellar and ionised gas kinematics are markedly different to each other, with an offset of $\sim$50\,\kms in the centre. However, \citet{Villforth15} note that the \oiii{} peaks do not coincide spatially with the X-ray sources, indicating that they may not be directly due to the two AGN, and suggest a bi-conical outflow as a likely alternative. A counter-rotating gas disc is also suggested, but disfavoured due to the ubiquity of outflows in AGN.

There are some important differences between 272831 and J1715: 272831 has much lower \oiii{} equivalent width and \oiii/H$\beta$ ratio, and the rotation curve of 272831 flattens at large radii. These differences favour a rotating disc description for the ionised gas in 272831, unlike J1715.

In summary, the IFS observations of 272831 do not provide any direct evidence for the existence of a binary SMBH system. Under one interpretation of the observations -- an ongoing merger of two galaxies, each providing one SMBH -- we can infer the presence of a second SMBH in the galaxy, although doing so rests on a number of unproven assumptions. In any case, the unusual kinematic signature that brought attention to this galaxy is not the result of a binary SMBH system, but is related to large-scale motions of the gas across the entire galaxy.

\subsection{9016800393}

The bulk of the ionised gas in 9016800393 shows rotational motion in the same sense as the stars, with a difference in the maximum velocities due to asymmetric drift. The exception to this ordered motion is in the centre, where two kinematic components are seen. The first component is narrow and has velocity consistent with that of the rotating gas and the stars. The emission-line flux ratios indicate that this component arises from the NLR of the AGN. However, the observed blueshift of the \oiii{} lines relative to the others indicates a more complex kinematic and ionisation structure than expected in a classic NLR.

The second component is broader (304\,\kms) and blueshifted by 230\,\kms{} along the line of sight. The \nii/H$\alpha$ ratio is similar to that of the NLR, but the \oiii/H$\beta$ ratio is higher, and the \sii/H$\alpha$ ratio is low. The broadened, blue-shifted gas suggests we are observing an outflow directed along or close to the line of sight. We note that the relatively strong \oiii{} emission rules out an AGN broad line region (BLR) as the source of the broader component, which in any case is significantly narrower than the $>$1000\,\kms{} dispersion typical of BLRs. The slight spatial extension of the broader component also favours a kpc-scale outflow rather than a BLR or kinematic feature close to the accretion disc.

It is likely that the blueshift of the narrow \oiii{} lines is a further consequence of the complex outflow structure. Comparison of the OSSY and two-component LZIFU measurements of the ionised gas velocity suggest the OSSY measurement was influenced by the strong \oiii{} lines. Hence, it appears that the identification of an offset between the stars and ionised gas was due to the outflow, rather than the true velocity of the AGN NLR.

AGN outflows are known to be ubiquitous, with a range of observed properties, and have been shown to produce velocity offsets like those seen here \citep{Veilleux94,Veilleux05}. For example, \citet{McElroy15} found complex emission features indicating winds in each of the 17 luminous Type-2 AGN they observed. Of their AGN, J103600, J103915, and J152637 are particularly close to 9016800393 in terms of their line ratios and velocity dispersions.

The SFR inferred from stellar populations in 9016800393 shows a recent burst, with a peak and subsequent decline within the last $10^8$\,yrs. Comparison to the H$\alpha$ SFR suggests this decline has continued: in the outer regions of the galaxy, the current SFR measured from H$\alpha$, $0.060\pm0.007$\,M$_\odot$\,yr$^{-1}$, is a factor of $\sim$10 lower than the SFR inferred from the youngest stellar populations. We speculate that the decline in SFR could be caused by negative feedback from the AGN outflow, but testing this scenario requires a larger sample of AGN and non-AGN.

The metallicity of the ionised gas in 9016800393 is typical for the galaxy's stellar mass, so unlike 272831 there is no evidence of an external origin for the gas.

Overall, the observations of 9016800393 strongly disfavour a binary SMBH interpretation, and instead point towards an outflow from a normal AGN. As noted by CG14 and others, such systems are an inevitable contaminant in samples of candidate binary SMBHs selected by kinematic offsets; spatially resolved spectroscopy provides a powerful test for these candidates.

\section{Conclusions}

We have examined in detail two galaxies that appeared to show kinematically offset AGN, using IFS data from the SAMI Galaxy Survey. The rich IFS data allow stronger conclusions to be drawn regarding the nature of these galaxies. In each case the IFS observations do not provide any direct evidence for the existence of a binary SMBH system.

In the first galaxy, 272831, there is a strong discrepancy between the kinematic properties of the ionised gas and those of the stars, suggesting a recent merger or gas accretion event. The stellar populations suggest that the star formation associated with the event began only 0.1\,Gyr ago. The ionisation properties in the centre of the galaxy are consistent with shock heating, which could be due to the AGN or simply the result of accreting gas onto the galaxy.

In the second galaxy, 9016800393, the AGN itself is kinematically consistent with the host galaxy. However, it is powering an outflow that appears blueshifted along the line of sight to the centre of the galaxy, producing the observed kinematic offset.

In both cases, we find no direct evidence for the existence of a second SMBH. In the first case we cannot rule out the possibility of a second SMBH, but the observed kinematic offset is not due to a binary system, and inferring the existence of the second SMBH would rely on a number of unproven assumptions.

\section*{Acknowledgements}

We are grateful to the anonymous referee for their report, which helped improve the quality and clarity of this work. We thank Luca Cortese, Ned Taylor, Madusha Gunawardhana and Julia Comerford for useful discussions and comments on a draft of this paper. We also thank Julia Comerford for early access to the CG14 sample of kinematically offset AGN.

The SAMI Galaxy Survey is based on observations made at the Anglo-Australian Telescope. The Sydney-AAO Multi-object Integral field spectrograph (SAMI) was developed jointly by the University of Sydney and the Australian Astronomical Observatory. The SAMI input catalogue is based on data taken from the Sloan Digital Sky Survey, the GAMA Survey and the VST ATLAS Survey. The SAMI Galaxy Survey is funded by the Australian Research Council Centre of Excellence for All-sky Astrophysics (CAASTRO), through project number CE110001020, and other participating institutions. The SAMI Galaxy Survey website is http://sami-survey.org/ .

JTA and ISK each acknowledge the award of a John Stocker Postdoctoral Fellowship from the Science and Industry Endowment Fund (Australia). SMC and MSO each acknowledge the support of the Australian Research Council through a Future Fellowship (FT100100457 and FT140100255).

This research made use of Astropy, a community-developed core Python package for Astronomy \citep{Astropy13}; the NASA/IPAC Extragalactic Database (NED), which is operated by the Jet Propulsion Laboratory, California Institute of Technology, under contract with the National Aeronautics and Space Administration; and NASA's Astrophysics Data System.

\bibliographystyle{mn2e}
\bibliography{offset_agn}{}

\end{document}